\documentclass[twocolumn,aps,amsmath,amssymb,nofootinbib,pre,floatfix]{revtex4}
\usepackage{amssymb,amsfonts,textcomp}
\usepackage{array}
\usepackage{supertabular}
\usepackage[normalem]{ulem}
\usepackage{hhline}
\usepackage{graphicx}
\usepackage{xcolor}
\usepackage{soul}
\usepackage[T1]{fontenc}

\newcommand{\stk}[1]{\ifmmode\text{\sout{\ensuremath{#1}}}\else\sout{#1}\fi}
\newcommand{\av}[1]{\langle {#1} \rangle}

\newcommand{\blacktxt}[1]{\textcolor{black}{#1}}

\newcommand{\be}{\begin{equation}}
\newcommand{\ee}{\end{equation}}

\begin{document}

\title{Synchronization transitions on connectome graphs 
with external force}

\author{G\'eza \'Odor (1), Istv\'an Papp(1), Shengfeng Deng(1) and Jeffrey Kelling (2,3)}
\affiliation{(1) Centre for Energy Research, Institute of Technical Physics
	and Materials Science, \\ P. O. Box 49, H-1525 Budapest, Hungary \\
	(2) Faculty of Natural Sciences, Chemnitz University of Technology, 09111 Chemnitz, Germany\\
	(3) Department of Information Services and Computing,
	Helmholtz-Zentrum Dresden-Rossendorf, P.O.Box 51 01 19, 01314 Dresden, Germany
}

\begin{abstract}

We investigate the synchronization transition of the
Shinomoto-Kuramoto model on networks of the fruit-fly
and two large human connectomes. This model contains a force term,
thus is capable of describing critical behavior in the presence of
external excitation. By numerical solution we determine the
crackling noise durations with and without thermal noise and 
show extended non-universal scaling tails characterized by
$2< \tau_t < 2.8$, in contrast with the Hopf transition of the
Kuramoto model, without the force $\tau_t=3.1(1)$. 
Comparing the phase and frequency order parameters we
find different transition points and fluctuations peaks as
in case of the Kuramoto model.
Using the local order parameter values we also determine
the Hurst (phase) and $\beta$ (frequency) exponents and compare
them with recent experimental results obtained by fMRI. We show
that these exponents, characterizing the auto-correlations are
smaller in the excited system than in the resting state and
exhibit module dependence.

\end{abstract}

\maketitle

\section{Introduction}

The critical brain hypothesis has been confirmed experimentally many times since the pioneering electrode experiments in~\cite{BP03}.
Power law (PL) distributed neuronal avalanches were shown in neuronal
recordings (spiking activity and local field potentials, LFPs) of neural
cultures in vitro~\cite{Mazzoni-2007,Pasquale-2008,Fried}, LFP
signals in vivo~\cite{Hahn-2010}, field potentials and functional magnetic
resonance imaging (fMRI) blood-oxygen-level-dependent (BOLD) signals in 
vivo~\cite{Shriki-2013,Tagliazucchi-2012}, voltage imaging in 
vivo~\cite{Scott-2014}, $10$--$100$ and single-unit or multi-unit 
spiking and calcium-imaging activity in 
vivo~\cite{Pris,Bellay-2015,Hahn-2017,Seshadri-2018}.
Furthermore, source reconstructed magneto- and electroencephalographic
recordings (MEG and EEG), characterizing the dynamics of ongoing 
cortical activity, have also shown nonuniveral PL scaling in neuronal 
long-range temporal correlations~\cite{brainexp,PalvanewGP}.
Optical methods, like light-sheet microscopy with GCaMP zebrafish 
larvae~\cite{ZfishC} or calcium imaging recordings of dissociated 
neuronal cultures~\cite{Yag} also show PL scaling. 

From a theoretical point of view the hypothesis is also very attractive as 
critical systems possess optimal computational capabilities as
well as provide efficient long range communications, memory and
sensitivity~\cite{ChialvBak1999_LearningMistakes, 
RChial2004_CriticalBrainNetworks,Chialv2006_OurSensesCritical,
Chialv2007_BrainNear,ChialvBalenzFraima2008_Brain:What,
FraimaBalenzFossChialv2009_Ising-likeDynamicsLarge-scale,
expert_self-similar_2011,FraimaChialv2012_WhatKindNoise,
Deco12,66,Senden16,MArep}.

Homogeneous critical systems exhibit universal scaling behavior and
many experiments claim indeed a mean-field class behavior of the
branching process~\cite{Plenz21soc,KC} generated by self-organized 
criticality~\cite{SOC}.
However, neural systems are very non-homogeneous, thus it is natural
to expect non-universal behavior, known in statistical physics 
within the field of quenched disordered models~\cite{Vojta2006b,CCrev}. 
Indeed some experiments~\cite{brainexp,PalvanewGP,Yag} show that 
the measured exponents are not universal, significantly different from
the mean-field class ones of the branching process.

Furthermore, external sources can move the system away from
criticality~\cite{Fosque-20} or tune it to other classes
like the isotropic percolation~\cite{Yag,PhysRevX.11.021059} 
or tricritical points~\cite{PhysRevE.106.054140}. 
More complex models than the two-state branching process, can also 
exhibit hybrid type of phase transitions, like threshold 
models~\cite{HPTcikk}, models with inhibitory 
nodes~\cite{PhysRevResearch.4.L042027} or models with oscillatory 
units~\cite{PhysRevResearch.3.023224}. 
Subsystems can also show different scaling behavior and may be 
within different distances from 
criticality~\cite{Morales2021.11.23.469734}.

For quenched disordered models it has recently been shown 
\cite{PhysRevE.106.044102,GProbcikk}, that even for weak time 
dependence the semi-critical, dynamical scaling,  which occurs in an 
extended control parameter region of criticality, 
in the so called Griffihts Phases~\cite{Griffiths} (GP), remains stable.
Furthermore, even when the network dimension is high, one does not 
find the usual mean-field behavior, but in the presence of modules a
GP~\cite{MM,HMNcikk,Cota_2018,HPTcikk,PhysRevX.11.021059,Buend_a_2022} 
or Griffihts effects ~\cite{Cota2016} and a different, sometimes
logarithmically slow scaling at the critical point ~\cite{Vojta2006b}.

The big advantage of critical universality is that more realistic models 
for the brain, like the integrate and fire models~\cite{Burkitt2006}, 
can also show the same criticality as simpler ones like in a 
recent work ~\cite{10.3389/fnsys.2020.580011}, which derives  Hopf 
bifurcation criticality or in a more experimental 
study~\cite{2013NatPh...9..582O} of neural cultures
agreement with isotropic percolation avalanche size distributions 
is obtained. 
But of course the directed percolation criticality~\cite{DPuni,DPuni2}, 
which occurs in branching processes~\cite{BP03} is the main 
example for the universality principle~\cite{odorbook}. 
Therefore, the study of simpler models, for which numerical analysis 
can be done are very useful for the brain science~\cite{MArep,CCrev}.

Recently threshold models and Kuramoto type of models have been analyzed
on different, available connectome networks and GP behavior was 
reported~\cite{CCdyncikk,GProbcikk,KurCC,KKIdeco,Flycikk,CCrev}.
This behavior is also called as frustrated 
synchronization~\cite{Frus,Frus-noise,FrusB} and has been analyzed 
within the framework of a Kuramoto like models, albeit
lacking quenched self-frequencies. 

From the experimental directions the different behavior in modules of 
brains of the mouse~\cite{Morales2021.11.23.469734}, by phenomenological
renormalization-group analysis of the spectrum of electrode spikes, and
humans~\cite{Ochab2022}, via Hurst and $\beta$ exponents analysis of fMRI;
quasi-critical (off-critical) scaling like behavior has been shown.
Here we attempt to model this using the Shinomoto--Kuramoto (SK) model
on connectomes of the fruit-fly (FF) and humans.
This is an extension of the Kuramoto model~\cite{kura}, which itself
does not have an external source, that can describe the resting state
critical behavior at the Hopf transition towards a model with a periodic
external driving force, thus may be appropriate to characterize 
criticality with an excitation~\cite{PhysRevResearch.3.023224}.

\section{Models and methods}

In this Section we introduce the synchronization model,
followed by an overview of different connectome graphs, 
on which we run the numerical analysis. Finally we discuss 
the method of local synchronization to dig into the 
details of the spatio-temporal simulations of these brain
systems.

\subsection{The Shinomoto--Kuramoto (SK) model}

We consider an extension of the Kuramoto model~\cite{kura} of 
interacting oscillators sitting at the nodes of a network, whose 
phases $\theta_j(t)$, $j=1,2\dots,N$ evolve according to the following 
set of dynamical equations
\begin{eqnarray}
\dot\theta_j(t) &=& \omega_j^0+K\sum_k W_{jk}\sin[\theta_k(t)-\theta_j(t)] \\
\nonumber
&+& F\sin(\theta_j(t)) + \epsilon\eta_j(t) \ .
\label{diffeq}
\end{eqnarray}
Here, $\omega_j^0$ is the so-called self-frequency of the $j$th 
oscillator, which is drawn from a Gaussian distribution with zero 
mean and unit variance. The summation is performed over 
adjacent nodes, coupled by the $W_{jk}$ matrix. 
Up to this point we have the classical Kuramoto model~\cite{kura}. 
In the Shinomoto extension~\cite{10.1143/PTP.75.1105}, we have a 
Gaussian annealed noise term $\eta_j(t)$, with an amplitude 
$\epsilon$, and to describe excitation, a site dependent periodic 
force term, proportional to a coupling $F$.

Sakaguchi~\cite{10.1143/PTP.79.39} was the first to study the 
periodically forced Kuramoto model. In numerical simulations, however, 
he found that the state of forced entrainment was not always 
attained: macroscopic fractions of the system self-synchronized at a 
different frequency from that of the drive, indicating that this 
sub-population had broken away and established its own collective rhythm. 
Analytically improvements were provided
in~\cite{doi:10.1063/1.2952447,doi:10.1063/1.2930766,doi:10.1063/1.3049136}
and found a rich phase space of the SK model.

Recently, in~\cite{PhysRevResearch.3.023224} the avalanche behavior 
of the full equation was investigated, albeit with site independent
self-frequencies $\omega_j^0 = \omega$. The authors explored the phase
diagram, besides the forceless Hopf transition a so-called saddle 
node invariant cycle (SNIC) and a hybrid type of bifurcation
was compered.
In a very recent publication~\cite{Buend_a_2022} this numerical
analysis has been continued on Erd\H{o}s--R\'enyi (ER) and hierarchical modular
networks, motivated by brain research. Considering quenched 
$\omega_j^0$-s with bi-modal frequency distributions the authors 
claim the emergence of Griffiths effects by the broadening of the
synchronization transition region.

Here we study the SK model using quenched $\omega_j^0$-s 
with and without annealed noise $\eta_j(t)$ on real connectomes.
In particular we test if the chaoticity, generated by the quenched 
$\omega_j^0$-s generates the same phase transition behavior and
avalanches as with the presence of the stochastic noise.
We measured the Kuramoto phase order parameter:
\begin{equation}\label{ordp}
z(t) = r(t) \exp{i \theta(t)} = 1 / N \sum_j \exp{[i \theta_j(t)}] \ ,
\end{equation}
by increasing the sampling time $\delta t = 0.01$.
Here $0 \le r(t) \le 1$ gauges the overall coherence and 
$\theta(t)$ is the average phase. 
The set of equations (\ref{diffeq}) was solved by the steppers 
Runge--Kutta-4 (RK4), for the noisy, or by the Bulrisch--Stoer~\cite{BS,BS2}
(BS) for the noiseless cases, because in the presence of noise 
the adaptive BS fails to work. Here and in earlier 
studies~\cite{KKIdeco} we found that the 
stronger stochastic noise makes the results non-reliable, while
application of other steppers slow down the numerical solution.
For the noisy cases we also tried the Euler--Maruyama solver~\cite{EM}, which
has a stronger mathematical foundation for stochastic differential equations.
This had to be restricted to testing purposes only, as this first-order
solver is orders of magnitude slower than the RK4 for the same precision.

We integrated the set of equations numerically for $10^3 - 10^4$ 
independent initial conditions, by different $\omega_j^0$-s and 
sample averages of the phases
\begin{equation}\label{KOP}
R(t) = \langle r(t)\rangle
\end{equation} 
and of the variance of the frequencies
\begin{equation}\label{FOP}
\Omega(t) = \frac{1}{N} \sum_{j=1}^N (\overline\omega(t)-\omega_j^2(t)) 
\end{equation}
were determined, where $N$ denotes the number of nodes.

In the steady state, which we determined by visual inspection of
$R(t)$ and $\Omega(t)$, we measured their half values and 
the standard deviations: $\sigma(R(t))$, $\sigma(\Omega(t))$ 
in order to locate the transition points. 
Note, that $\sigma(R(t))$ is just a version
of the SK order parameter employed by~\cite{PhysRevE.100.062416}
for discrete version of oscillatory models.
In case of the Kuramoto equation the fluctuations of both order 
parameters show a peak, albeit at different $K'_c$ (for phases)
and $K_c$ (for frequency) values in the case of the KKI-18 
connectome. For graph dimensions $3 < d < 4$ a crossover 
transition is expected for $R$ and phase transition for $\Omega$.
In the case of the FF, having $d > 5$ we found $K_c \simeq K'_c$, 
which is expected for real phase transitions at large sizes, 
where both order parameters converge to a finite value in 
the infinite size limit~\cite{Flycikk}.

\subsection{Connectome graphs }

The connectome is defined as the structural network of neural
connections in the brain~\cite{sporns_human_2005}.
For the fruit-fly connectome, we used the hemibrain data-set (v1.0.1) 
from~\cite{down-hemibrain1.0.1},
which has $N_{FF}=21\,662$ nodes and $L_{FF}=3\,413\,160$ edges, out of which
the largest single connected component contains $N=21\,615$ and
$L=3\,410\,247$ directed and weighted edges.
The number of incoming edges varies between $1$ and $2708$.
The weights are integer numbers, varying between $1$ and $4299$.
The average node degree is
$\langle k\rangle= 315.129$ (for the in-degrees it is: $157.6$),
while the average weighted degree is $\langle w\rangle= 628$.
The adjacency matrix, visualized in ~\cite{Flycikk} where one can
see a rather homogeneous, almost structureless network,
however it is not random.
For example, the degree distribution is much wider than that of a random
ER graph and exhibits a fat tail.
The analysis in~\cite{Flycikk} found a weight distribution $p(w)$
with a heavy tail, assuming a PL form, an exponent $-2.9(2)$ could be 
fitted for the $w > 100$ region.

The human brain has $\approx 10^{11}$ neurons, which current imaging
techniques cannot comprehensively resolve at the scale of single neurons. 
We used graphs on the coarse-grained, level with $\approx 10^{6}$ nodes 
obtained by diffusion tensor imaging~\cite{landman_multi-parametric_2011}.
This method has generally been found to be in good agreement with 
ground-truth data from histological tract 
tracing~\cite{delettre_comparison_2019}.
Inferred networks of structural connections were made available by
the Open Connectome Project and previously analyzed 
by~\cite{gastner_topology_2016}.
These graphs are symmetric, weighted networks, where the weights
measure the number of fiber tracts between nodes.
The network topology study found a certain level of universality in the
topological features of the ten large human connectomes investigated:
degree distributions, graph dimensions, clustering and small world 
coefficients.
These can be observed in Tables 3 and 4 of~\cite{gastner_topology_2016}.
Therefore, two networks, called KKI-18, and KKI-113 were selected
to be the representatives in further studies.
The graphs, downloaded in 2015 from the Open Connectome project
repository~\cite{OCP}, were generated via the MIGRAINE 
pipeline~\cite{MIG}, publicly available from~\cite{m2g}.
KKI-18 comprises a large component with $N = 804\,092$ nodes connected via
$41\,523\,908$ undirected edges and several small disconnected sub-components,
which were ignored in the modeling. Similarly, the extracted largest 
connected component of KKI-113 contains $799\,133$ nodes connected by 
$48\,096\,500$ undirected and weighted edges.
The large number of nodes is because of other parcellations closer to
voxel resolution being used. For instance, there are approximately 
1.8 million voxels in the brain mask of a 1\,mm resolution 
standard-aligned MRI.
The graphs exhibit a hierarchical modular structure, because they are constructed
from cerebral regions of the Desikan--Killany--Tourville parcellations,
which is standard in neuroimaging~\cite{DESIKAN2006968,10.3389/fnins.2012.00171}
providing (at least) two different scales.

The modularity quotient of a network is defined by \cite{Newman2006-bw}
\begin{equation}
Q=\frac{1}{N\av{k}}\sum\limits_{ij}\left(A_{ij}-
\frac{k_ik_j}{N\av{k}}\right)\delta(g_i,g_j),
\end{equation}
the maximum of this value characterizes how modular a network is, where 
$A_{ij}$ is the adjacency matrix, $k_i$, $k_j$ are the node degrees of 
$i$ and $j$ and $\delta(g_i,g_j)$ is $1$ when nodes $i$ and $j$ were found to 
be in the same community, or $0$ otherwise. 
However, this value is not independent of the community detection method. 
If our detection method produces lower modularity than the maximum achieved, 
it means our algorithm has fallen behind others. 
Community detection algorithms based on modularity optimization will get 
the closest to the actual modular properties of the network. 
We calculated the modularity using community structures detected by the 
Louvain method \cite{Blondel2008}, the results for each network were: 
$Q_{FF} \approx 0.631, Q_{KKI-18} \approx 0.913, Q_{KKI-113} \approx 0.915$.
The FF is a small-world network, according to the definition of
the coefficient~\cite{humphries_network_2008}:
\begin{equation}
\sigma^W = \frac{C^W/C_r}{L/L_r} \ ,
\label{swcoef}
\end{equation}
because $\sigma_{FF} = 9.5$ is much larger than unity.
Here $C^W$ denotes the Watts clustering coefficient,
and $L$ the average path length.
$C_r$ and $L_r$ are the reference values of
random networks with the same sizes and average degrees.
The same is true for the human connectomes, as their $\sigma^W$
is in the range between $400$ and $1000$ 
~\cite{gastner_topology_2016}.

The effective graph (topological) dimension, obtained by the 
breadth-first search algorithm is $d=5.4(5)$. This is
defined by $N(r) \sim r^d$, where the number of
nodes $N(r)$ with chemical distance $r$ or less from the seeds 
are counted and averages are calculated over the many trials. 
For the Open Connectome data, power-law fits in the range $1\leq r
\leq 5$ suggest topological dimensions between
$d=3$ and $d=4$~\cite{gastner_topology_2016}.

As these structural connectome graphs exhibit heavy-tailed
weight distributions, probably as a result of learning,
there exist hubs, which could fully determine the behavior
of neighboring nodes and suppress the occurrence of 
critical behavior in the models~\cite{CCdyncikk}.
In reality, on top of the structural weights, there exist
inhibition/excitation mechanisms, which control the dynamics 
of the neural system and provide a local homeostasis.
As we do not know the details of these mechanisms, in earlier 
studies~\cite{CCdyncikk,GProbcikk,KurCC,KKIdeco,Flycikk,CCrev}, 
the weight normalization scheme
\begin{equation}
W'_{jk} = W_{jk}/\sum_k W_{jk}
\end{equation}
was applied to achieve this artificially. This way we equalize 
the sensitivity of nodes to the incoming excitation.
We do the same in the simulations presented here.

\subsection{Analysis of the local synchronization}

As the connectomes are very heterogeneous, built up from modules
we also measured the local Kuramoto order parameter $R_i(t)$, defined 
as the partial sum of phases for the neighbors of node $i$
\begin{equation}\label{LCO}
R_i(t)= \frac{1}{N_{\mathrm{i.neigh}}}
\left|\sum_j^{N_{\mathrm{i.neigh}}}  A_{ij} e^{i
	\theta_j(t)}\right|  \ ,
\end{equation}
and the local $\Omega_i(t)$ defined as
\begin{equation}\label{LFO}
\Omega_i(t) = \frac{1}{N_{\mathrm{i.neigh}}}
\left|\sum_j^{N_{\mathrm{i.neigh}}} 
(\overline\omega(t) - \omega_j(t))^2 \right|  \ .
\end{equation}
The local Kuramoto order parameter was initially suggested 
by~\cite{restrepo2005,schroder2017} to quantify the local
synchronization of nodes, which allows us to follow the synchronization
process by mapping the solutions on the connectome graphs.

The necessity of storing the states of the system at each time
step requires large amount of hard drive storage. Thus we analyzed 
the local order parameters in a time period of $50$ time-steps as 
stop time with time increment of $dt' = 0.1$ , in the steady state.  
To study it in more detail we also separated the networks into communities. 
Although, these communities should be separated according to anatomical 
and/or functional properties \cite{SANCHEZRODRIGUEZ2021117431}, we chose 
as a first approximation a community detection method based on global 
optimization of the modularity \cite{Blondel2008}. 
This method yielded 9 modules in FF network, 130 communities in KKI-113 
and 134 modules in the giant component of KKI-18. For detecting community 
structure that is closer to the real anatomical functional communities 
just by using the network topology, one might require other algorithms, 
which analyze the network with more depth, or even using fuzzy clustering 
methods \cite{Deritei_2014,PhysRevE.95.022306}.  

We studied the long-term persistence of the local order parameters 
with the Hurst and $\beta$ exponents. 
The Hurst exponent measures the degree of self-similarity of a time series, 
based on the assumption of an Ornstein--Ulenbeck process, that the 
measured values will go back to its average in just a few time-steps. 
The Hurst exponent is defined as follows:
\begin{equation}\label{LHO}
\mathbb{E} \left [ \frac{Z(n)}{S(n)} \right] = Cn^H,
\end{equation}
where $\mathbb{E}$ is the expectation value of the rescaled range 
$Z/S$ and $Z(n)$ is the cumulative deviate of the series until
the first number of $n$ data points ($n = (t_{max} - t_0 ) / dt'$), 
while $S(n)$ is the sum of the standard deviations until that point.
We averaged the first local parameter values within the 
communities and calculated the Hurst exponent over the $n$ points 
in the time period $t$, where
$S_j(n) = \sum_{i}^{M_{j,comm}} R_i(t)$ are community averages and
$M_{j,comm}$ is the number of nodes in the community.
We calculated the Hurst exponents for all communities. 

Similarly the power spectral scaling exponent, $\beta$, is used for 
quantifying long range correlations in time series. The power spectral 
density is the modulus of the Fourier transform, if the spectrum of 
the process satisfies a power-law scaling relation: 
\begin{equation}\label{BETA}
S(f) = \left | \sum_{j=0}^{N} \Omega_j(t) e^{-2\pi i f_j/N} 
\right|^2 \approx 1/f^{\beta},
\end{equation} 
where $f_j = \sum_{j}^{M_{i,comm}} \Omega_i(t)$ and $\beta$ must be 
obtained by using a linear fit to the logarithmic axes of the Fourier 
transform periodigram~\cite{Ochab2022}.

\section{Force driven synchronization transition}

First we determined the synchronization transition behavior
of the Shinomoto--Kuramoto model on different connectomes by
calculating the global order parameters $R$ and $\Omega$ as well
as their fluctuations as the function of the force control
parameter, which mimics the external excitation of the system.
After that we measured the crackling noise distributions
within the neighborhood of these transitions  

\subsection{Global order parameters}

\begin{figure}[!htbp]
        \centering
        \includegraphics[width=0.46\textwidth]{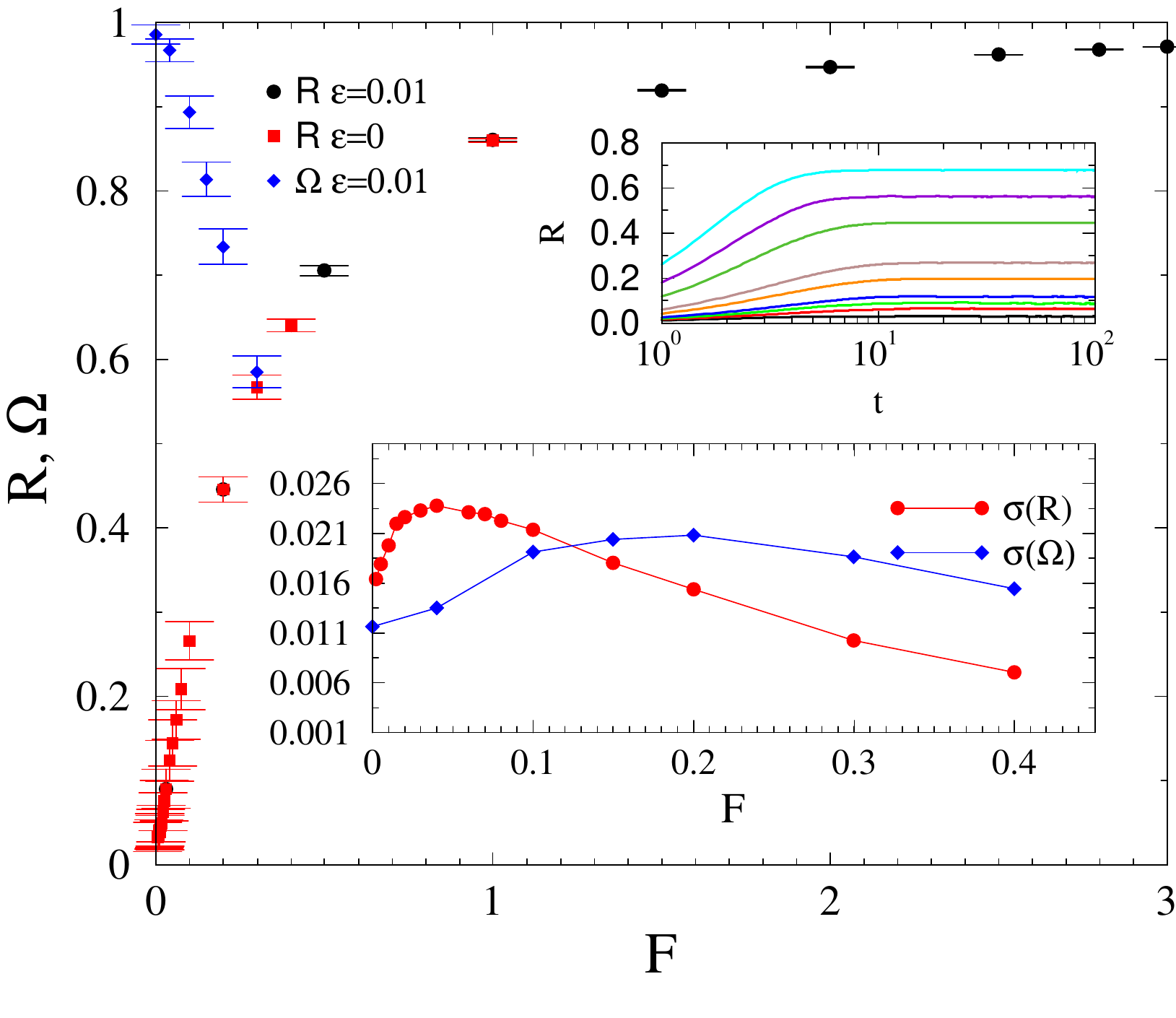}
        \caption{Order parameter dependence on $F$ for the fruit-fly
        connectome for the noisy (black bullet) and the noiseless 
        (red boxes) cases at $K=1.3$. The blue diamonds show the
        steady-state $\Omega$ values with noise.
        Lower inset: Variances of $R$ and $\Omega$ for the noisy case.
        Upper inset: Time dependence of the noisy $R(t)$, for $F=$ 
        0, 0.02, 0.03, 0.04, 0.07, 0.1, 0.2, 0.3, 0.4 (bottom to top 
        curves). \label{betaflyF_l1.3}}
\end{figure}

We started the numerical analysis of SK on the fruit-fly connectome 
at the global coupling value $K=1.3$, which was found to be 
asynchronous without a force in~\cite{Flycikk}. 
For each $F$ value we determined the steady state by following 
the evolution of the control parameters
starting from random initial $\theta$-s via visual inspection.
Averaging was done over many independent samples, corresponding
to different initial $\omega_j$ self-frequencies.
The transient regimes were short, in the range of 10-100 time 
steps and we could not see PL growth as in case of the Hopf 
transition of the Kuramoto model. But the Kuramoto order parameter 
curves exhibit $R(t) \propto  \ln(t)^{x(K)}$ type of growth 
(see upper inset of Fig.~\ref{betaflyF_l1.3}), as in case of 
activated scaling in disordered systems~\cite{Vojta2006b}.

To locate the transition we plotted the steady state values of
$R$ and $\Omega$ and their fluctuations on Fig.~\ref{betaflyF_l1.3}. 
The half values provide estimates: $F'_c \simeq 0.22$ and 
$F_c \simeq 0.35$.
One can see smooth fluctuation peaks of $\sigma(R)$ at 
$F' \simeq 0.04$ and of $\sigma(\Omega)$ at $F \simeq 0.2$.
Thus, the two different order parameters seem to exhibit 
different synchronization points. The frequency fluctuation peak
agrees roughly with $F'_c \simeq 0.22$, but the phase
fluctuation peak occurs at a much lower value. 
This, in contrast with the Hopf transition of FF and the random
network, where fluctuation peaks were roughly the same position,
where we knew that the dimension is $d > 4$.

As $\sigma(R)$ is also called SK order parameter, which
characterizes the transition in excitable systems, its
approach to zero as $F$ increases agrees with the SNIC
transition result of~\cite{PhysRevResearch.3.023224}, albeit
that was obtained in the synchronous phase. We have
also run SK in the synchronous phase of FF, using $K = 2$, 
and we found similar results as in the asynchronous phase.

Results with and without a small noise with amplitude $\epsilon=0.01$ 
did not show observable differences, so the chaotic noise from the
quenched disorder is capable to compete with the ordering effect
of the force.

\begin{figure}[!htbp]
        \centering
        \includegraphics[width=0.46\textwidth]{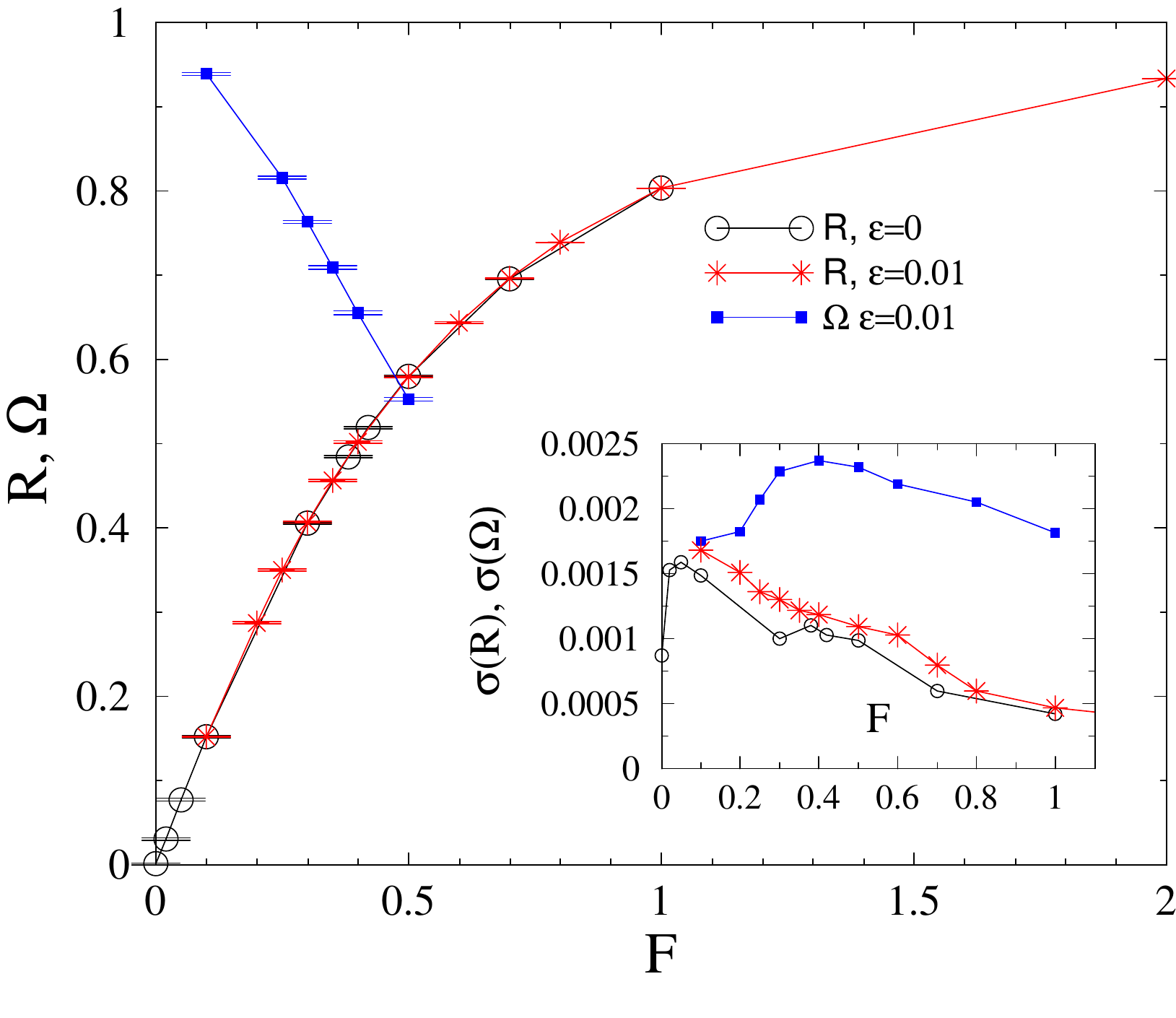}
        \caption{Order parameter dependence on $F$ for the 
        KKI-113 for the noisy and the noiseless cases at $K=1$.
        Inset: Variances of $R$ and $\Omega$ for the noisy case.
        \label{beta113F_l1}}
\end{figure}

As the next step we performed the same analysis of the human connectomes
at $K=1$, which is in the asynchronous phase without a force
~\cite{KurCC}.
Figure~\ref{beta113F_l1} shows the steady state values both for
$R$ and $\Omega$ in case of $K=1$ for KKI-113. 
Again the annealed noise does not modify the results and seems 
to be unnecessary to produce a synchronization transition. 
We estimated: $F'_c \simeq 0.4$ and $F_c \simeq 0.55$ by the half values
or $R$ and $\Omega$ respectively.
The fluctuation peaks of the two order parameters are again far away 
from each other: $F' \simeq 0.05$ versus $F \simeq 0.4$. Again the
fluctuation peak of $\Omega$ is close to $F'_c \simeq 0.4$.

\begin{figure}[!htbp]
        \centering
        \includegraphics[width=0.46\textwidth]{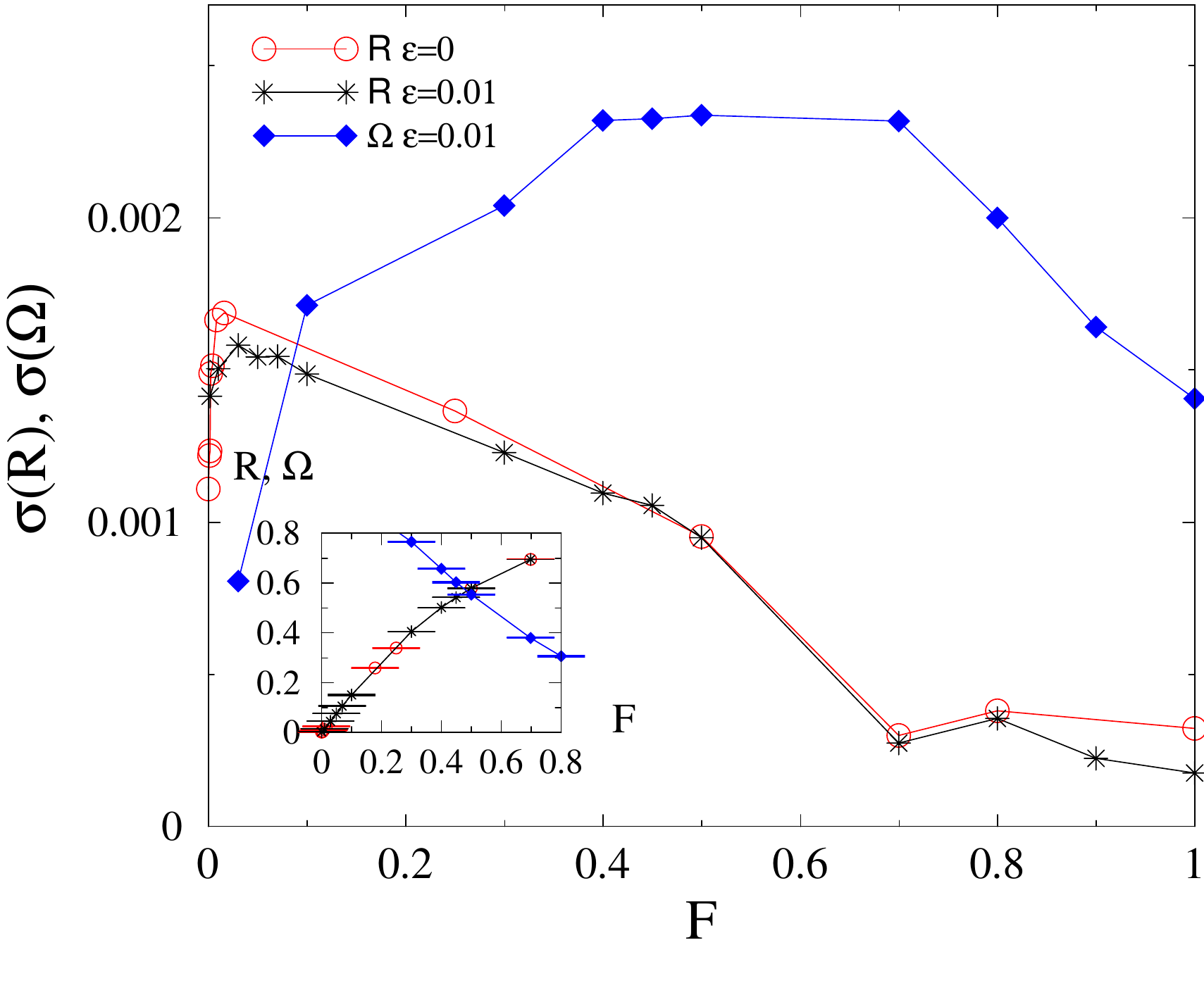}
        \caption{Fluctuations of $R$ and $\Omega$ as the function of 
        	$F$ for the KKI-18, for the noisy and the noiseless cases at $K=1$.
        	Inset: Order parameters for the noisy and noiseless cases..
        \label{sigma-ocp-kurl1}}
\end{figure}

For the connectome KKI-18 we enlarged the fluctuation peak
results on Figure~\ref{sigma-ocp-kurl1}. The smeared 
synchronization 'peaks' happen at similar values as for
KKI-113: $F' \simeq 0.05$ and $F \simeq 0.5$ within numerical 
precision. The transition points, estimated by the half values
of $R$ is $F'c \simeq 0.4$ and of $\Omega$ is $F_c \simeq 0.55$.
Again, the $\sigma(R)$ peaks are much lower than the 
other transition point estimates.

\subsection{Avalanche durations}

We investigated avalanches similarly to the local field potential
experiments and as it was done in simulations of spike-like 
events~\cite{PhysRevResearch.3.023224}. However, we did not 
threshold individual variables $\theta(t)_i$, but the global 
order parameter $R(t)$, which is a sum of them. This has the
advantage of a much faster algorithm, allowing to consider
larger statistics and the lack of ambiguity in avalanche definitions
~\cite{touboul2010can,villegas2019time,dalla2019modeling}. 
The disadvantage is that spatially independent avalanches
overlapping in time accidentally may be unified, thus the duration
times can be larger and we do not have information on the
spatio-temporal sizes, thus on the exponent $\tau$. Still, we
think that investigating this coarse-grained description of 
avalanches, which has also been measured in experiments, as a kind
of crackling noise~\cite{Sethna_2001} in the case of zebrafish 
larvae~\cite{ZfishC}, describes a possible critical behavior. 
Results of local characterization of the synchronization will be shown 
in Sects.~\ref{LCO}, \ref{HB}, \ref{HBnF}.

As in~\cite{PhysRevResearch.3.023224} here we also found that the 
choice of threshold $T(F)$ value did not change the scaling 
behavior of the duration distributions if it was chosen within the 
fluctuation range $R_\mathrm{min} < T(F) < R_\mathrm{max}$ 
corresponding to $F$, that was determined numerically after 
several runs on different 
initial conditions. For thresholds we used the mean 
value of $R(t)$, obtained in the steady state by sample and
time averaging up to $t_{max}=10^4$.
By the integration we used uniform random distributions 
$\theta_i(0) \in (0,2\pi)$ and the initial frequencies were set to 
be $\dot{\theta_i}(0)=\omega_i^0$. Following measurements of
the avalanche duration $\Delta(t)= t_{i}-t_{i'}$, defined between 
subsequent crossing of an up event:  $R(t_{i}) > T$  and a 
down one: $R(t_i') < T$, we applied a histogramming to determine 
the probability distributions $p(\Delta(t))$.

\begin{figure}[!htbp]
        \centering
        \includegraphics[width=0.45\textwidth]{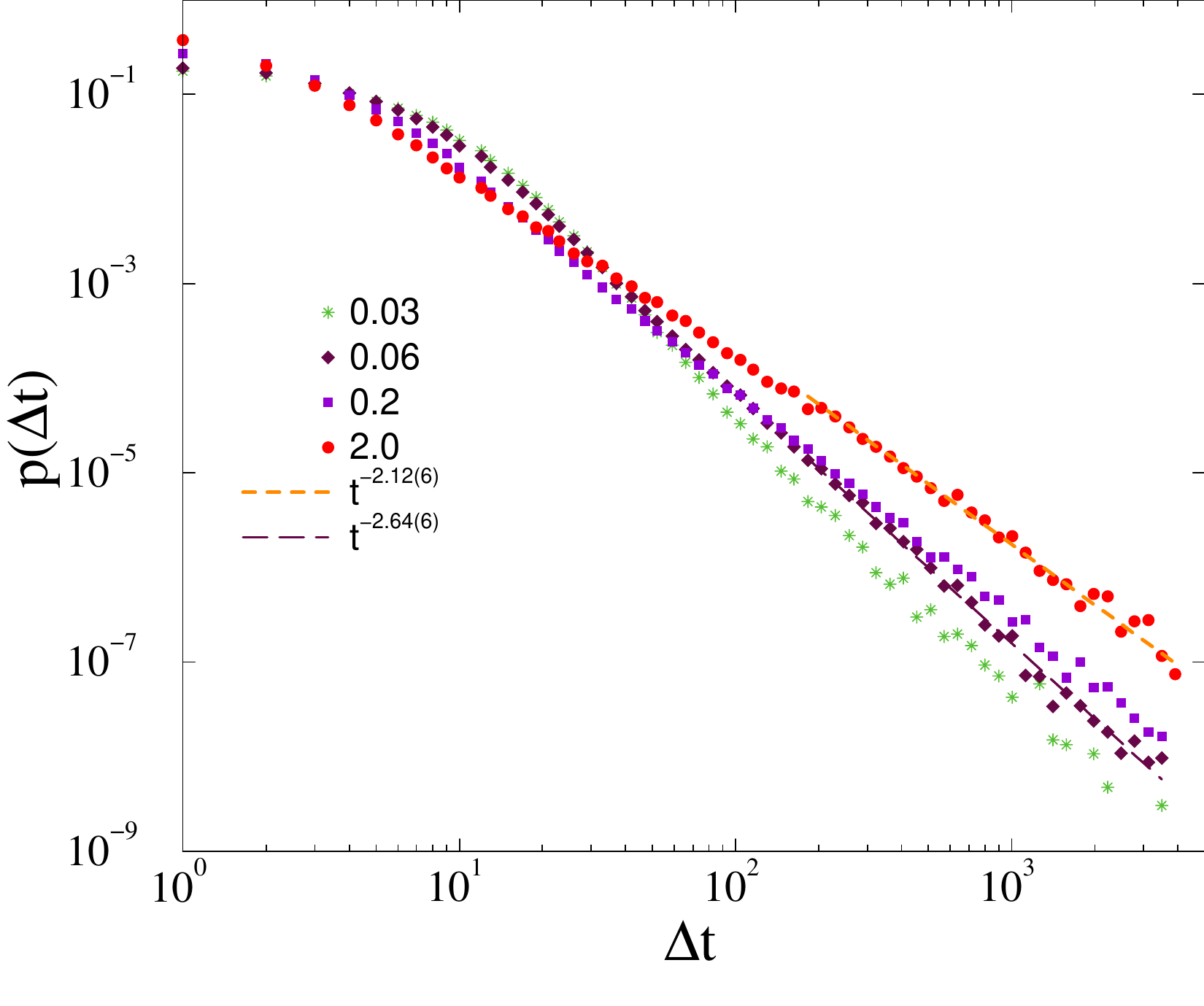}
        \caption{Avalanche duration distributions on the fruit-fly
        connectome for different forces, shown by the legends 
        and at $K=1.3$, $\epsilon=0.01$. Dashed lines are PL fits
        for $\Delta t > 100$.
        \label{elo-o.flyF_l1.3_N0.01}}
\end{figure}

Fig.~\ref{elo-o.flyF_l1.3_N0.01} shows the pdf-s $p(\Delta(t))$
results for the fruit-fly in case of $K=1.3$, $\epsilon=0.01$ 
and different forces. We can see $F$ dependent extended PL tails, 
with continuously changing exponents: $2.1 < \alpha < 2.8$,
which are somewhat smaller, but close to the experimental values 
for the zebrafish: $\alpha=3.0(1)$~\cite{ZfishC}.

\begin{figure}[!htbp]
        \centering
        \includegraphics[width=0.44\textwidth]{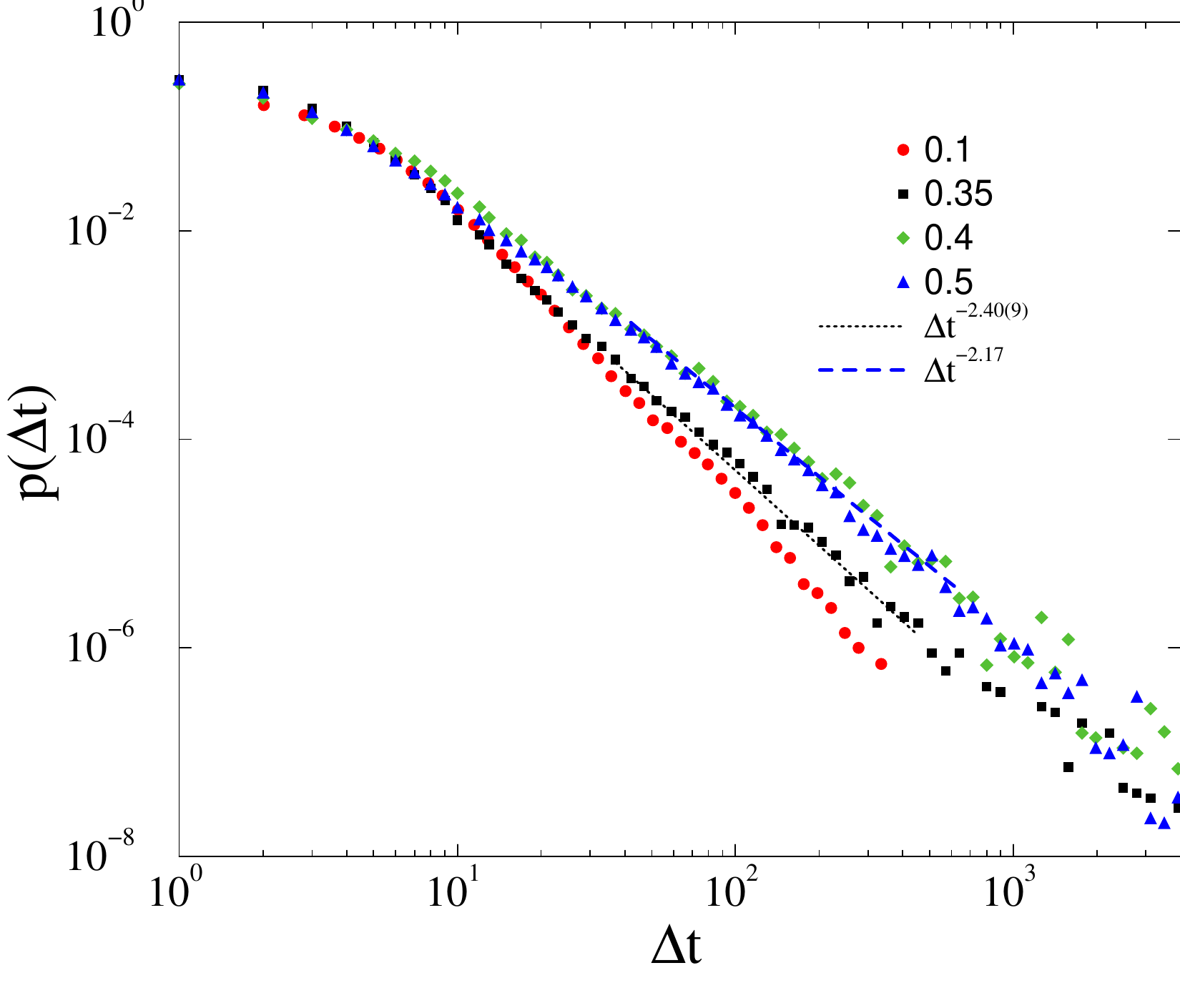}
        \caption{Avalanche duration distributions on the KKI-113
        connectome for different forces, shown by the legends and 
        at $K=1$, $\epsilon=0.01$. Dashed lines are PL fits
        for $\Delta t > 20$.
        \label{elo-o.113-kurl1_F_N0.01}}
\end{figure}

\blacktxt{Similar results are obtained in case of the two human connectomes
as shown on Figs.\ref{elo-o.flyF_l1.3_N0.01},\ref{elo-o.113-kurl1_F_N0.01}.
Furthermore, the results do not change without the additive noise,
or in case of a force in the synchronized phase (see graphs in the
Appendix)}.

\begin{figure}[!htbp]
	\centering
	\includegraphics[width=0.46\textwidth]{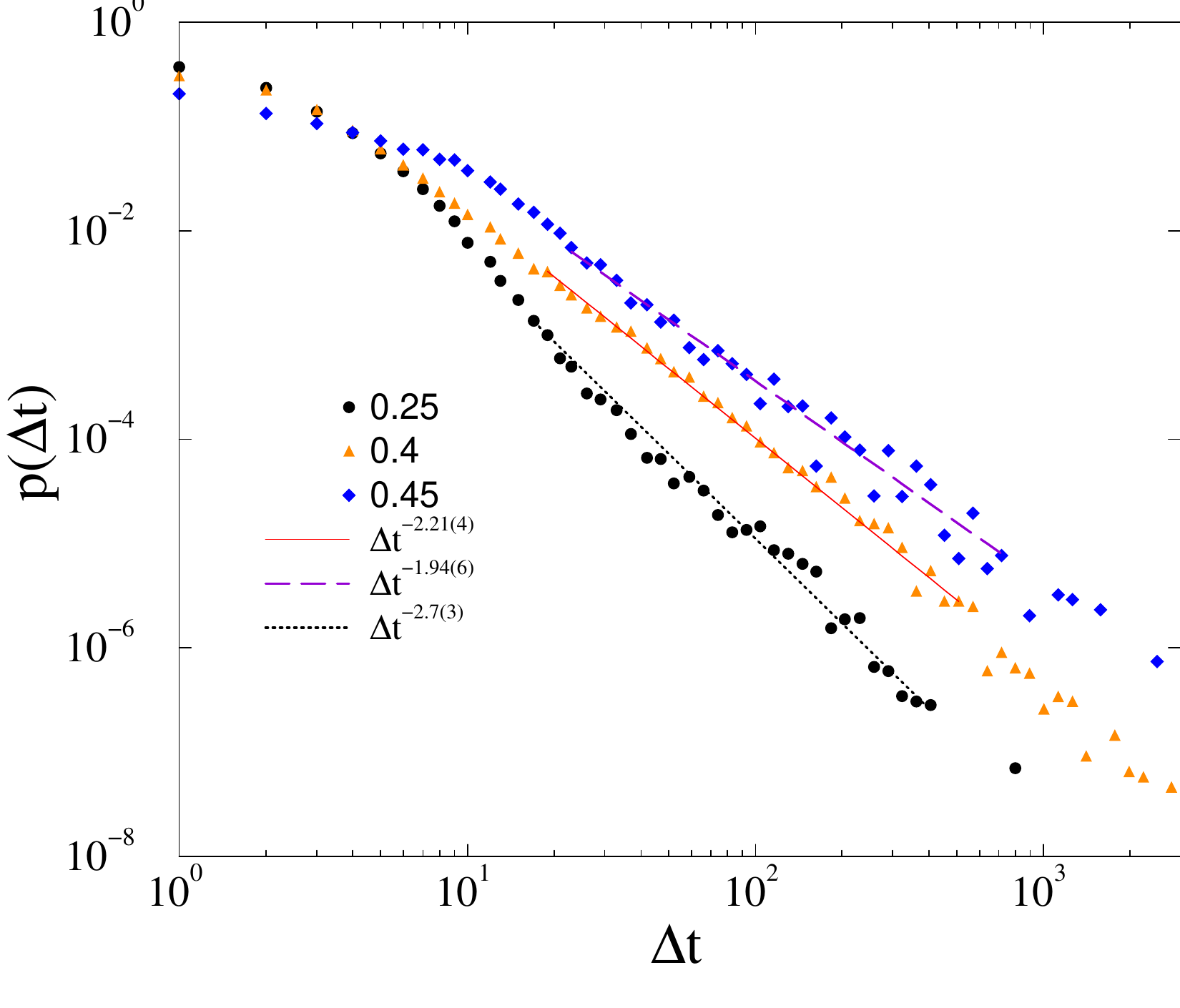}
	\caption{Avalanche duration distributions on the KKI-18
		connectome for different forces, shown by the legends 
                and at $K=1$, $\epsilon=0.01$. Dashed lines are PL 
                fits for $\Delta t > 20$.
		\label{elo-o.ocp-kurl1_F_N0.01}}
\end{figure}

\subsection{Local order parameters snapshots}\label{sec:LCO}

\begin{figure}[!htbp]
	\centering
	\includegraphics[width=0.46\textwidth]{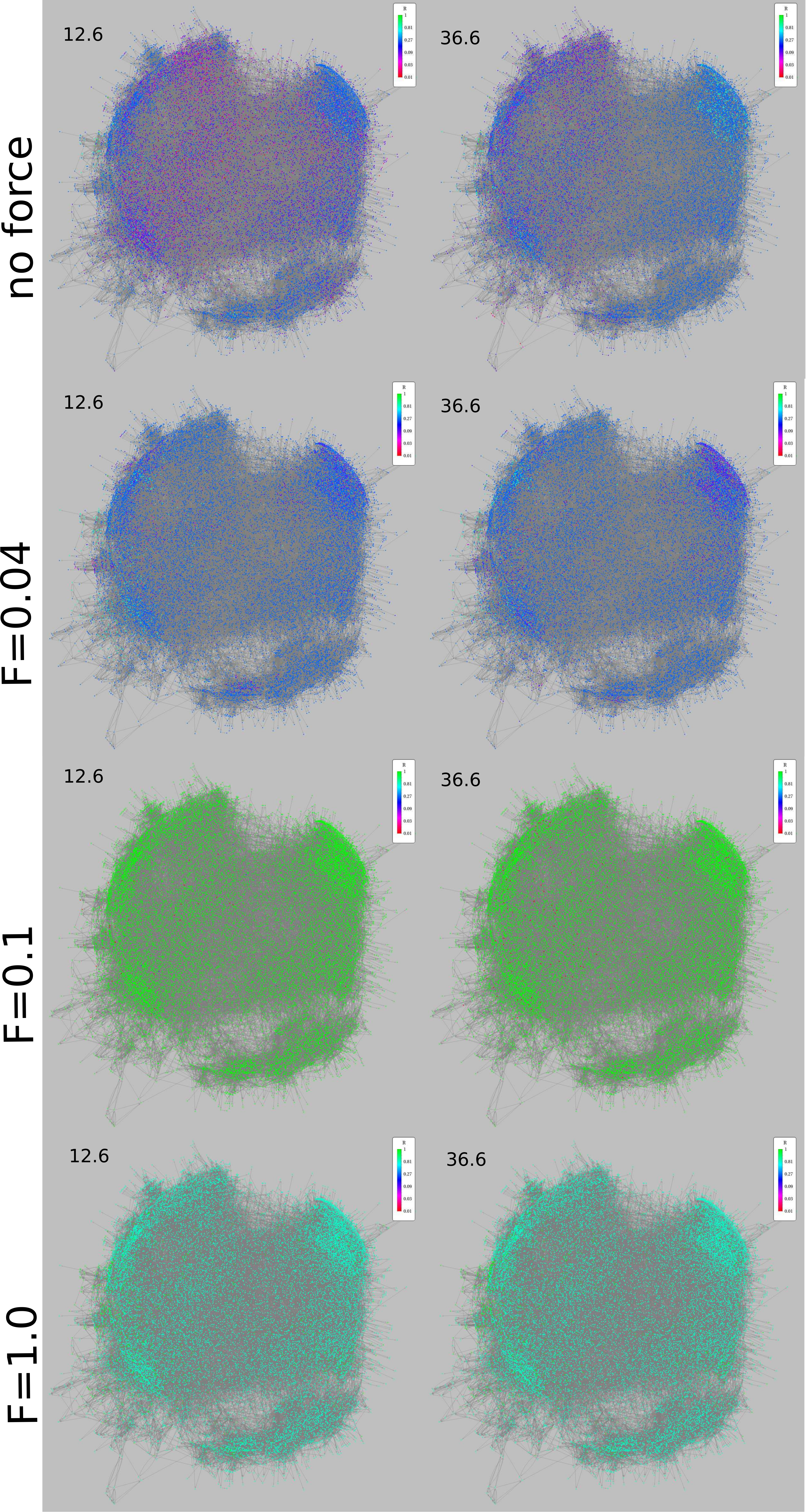}
	\caption{Here we see the evolution of the local order parameters 
		of a sub-graph of the fruit-fly connectome at different time steps:
		$t=12.6, 36.6$. The upper row shows $R_i$ map without a force,
		the lowest one with $F=1.0$.
		\label{snapshot}}
\end{figure}

We have plotted with Wolfram Mathematica
\cite{reference.wolfram_2022_graph} snapshots of the local 
order parameters of the FF at different force values in increasing 
order for the average local parameters (see Fig.~\ref{snapshot}).
The giant component of the graph was plotted with $21 \, 615$ nodes, 
however with a very few $75\,657$ edges for better visualization,
where we sorted the links of each 
node by their weights in a decreasing manner and then randomly chose the first 
$n_r$ links, where $n_r$ is a random integer between 1 and $n_m=6$.
Since the graph is a modular small-world graph, this kind of representation 
can be a close visual representation of the actual network. 
The color coding on the figure is a logarithmic ($\log_3$) binned 
scale between $0$ and $1$ (0.01, 0.03, 0.09, 0.27, 0.81, 1.)  
representing the $R_i$ values of each node at time step, indicated on 
the top left of each figure.

Top row plots are results without force, second row at $F=0.04$, third 
row at $F=0.1$ and last row is at $F=1.0$. 
Similarly to the $\beta$ exponent's case, 
we notice that the average local parameter $R$ is not increasing 
linearly with the force at the same time-step. 
There is a maximum around $0.1$,
thus it does not have a linear correlation with the force.
Without force the steady state has more fluctuations and the 
communities are more observable through visualization. 
By increasing the force every node comes into the same local state.

\subsection{Hurst and $\beta$  exponent results}\label{HB}

The $H$ and $\beta$ exponents measure the self-similarity 
of a time series, when power-law behavior 
(\ref{LHO}), (\ref{BETA}) can be observed. 
$H$ and $\beta$ values lower than $0.5$ describe 
anti-correlated signals. 
On the other hand, values between $0.5$ and $1$ mean signals 
with long range correlations in time.

\begin{figure}[!htbp]
   \centering
   \includegraphics[width=0.49\textwidth]{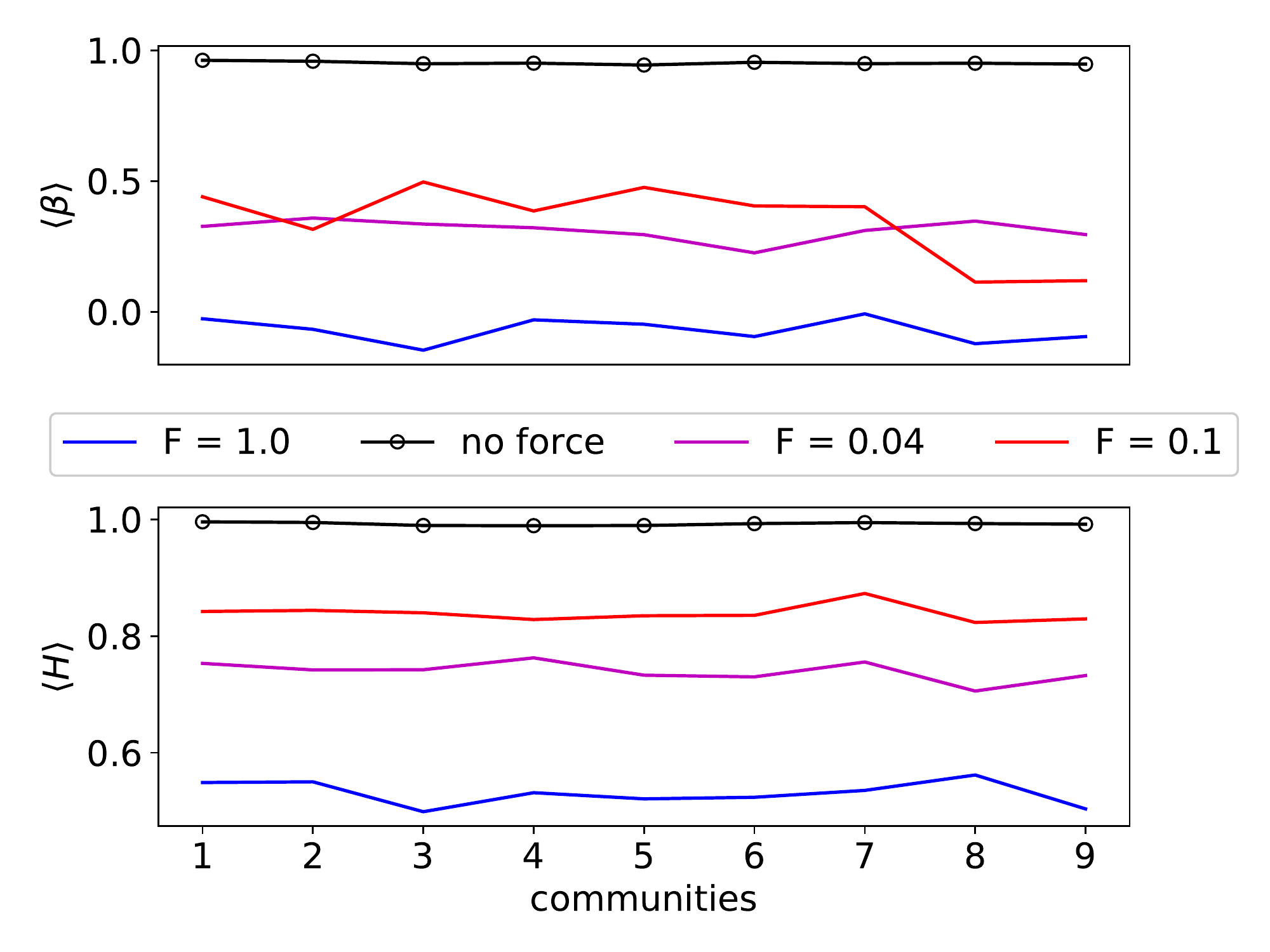}
   \caption{Hurst and beta exponents of all fruit-fly connectome communities.
   In the forceless case at the critical Hopf transition coupling, 
   the $H$ exponent is the largest for every community. 
   With forces these values drop for each community. This shows a 
   resemblance with the rest and non-rest studies of different 
   brain areas in \cite{Ochab2022}, showing $\langle H \rangle \approx 1.0$ 
   at resting state and $\langle H \rangle \approx 0.7$ at task driven states.
   \label{fruitfly-hurst}}
\end{figure}

First, we separated the communities in all FF, KK-18, KKI-113 connectomes
with the Louvain modularity optimizing algorithm. 
Then, we calculated the $H$ and $\beta$ exponents for each community 
for the local parameters. In case of the FF the results 
(see Fig.~\ref{fruitfly-hurst}
with force could similarly be differentiated from the results without 
force as in the ~\cite{Ochab2022} experiments with rest and task 
driven measures. Simulation results without force seem to have longer 
correlations in time, resembling to the fMRI measurements at 
the rest phase.

\begin{figure}[!htbp]
	\centering
	\includegraphics[width=0.49\textwidth]{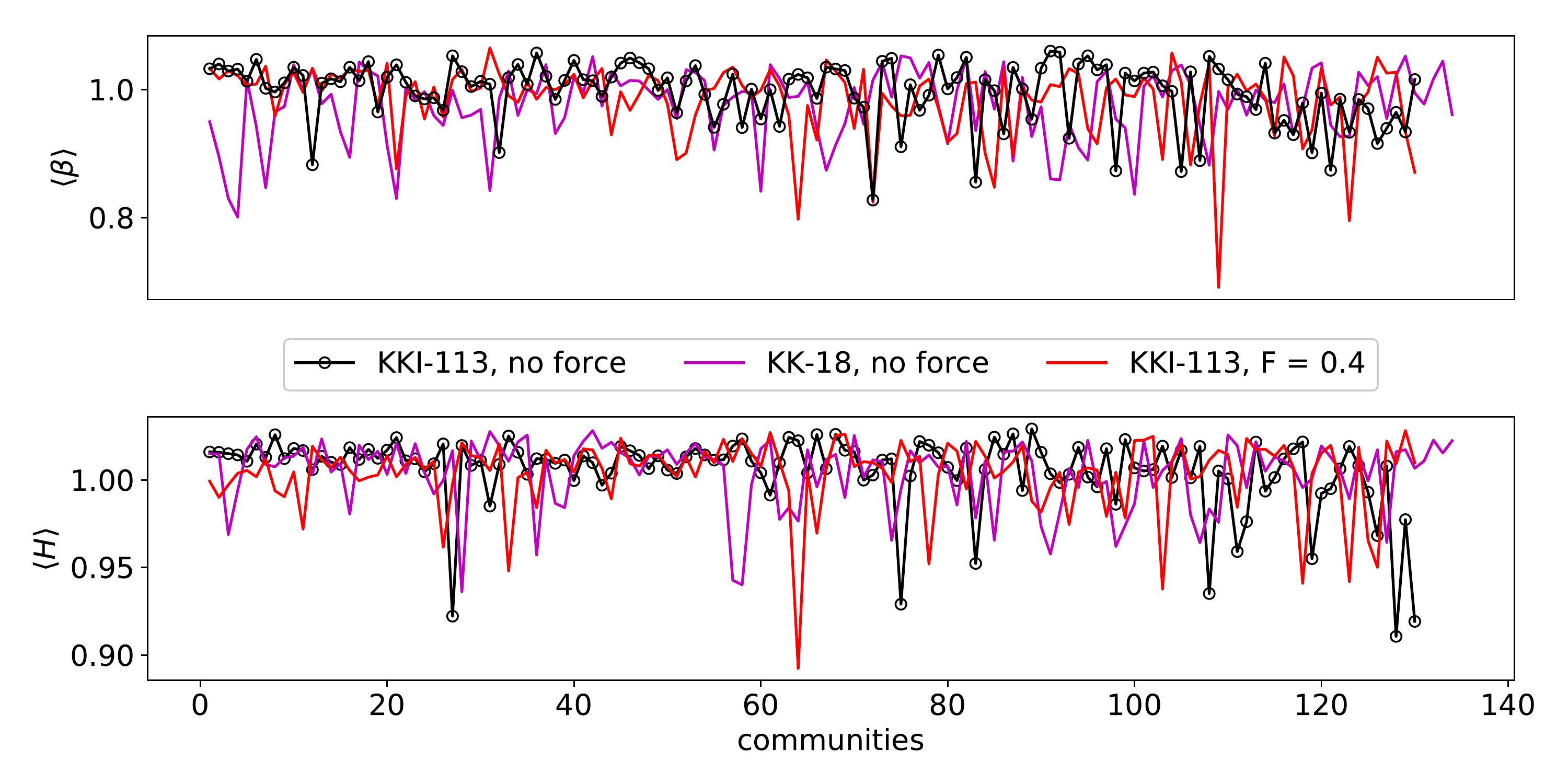}
	\caption{Hurst and $\beta$ exponents of all human connectomes' 
         communities. KKI-113 is presented with and without force terms and KK-18 without the force terms. 
	\label{human-hurst}}
\end{figure}

The same conclusion however cannot be found in the case of the human 
connectomes (see Fig.~\ref{human-hurst}).
It appears that even with a relatively high force the exponents remain 
close to each other and close to those of the ``rest" phase. 
In the case of FF higher force led to less ``rest" 
in the system resembling more like task driven behaviour. 

\section{Hopf synchronization transition without force}

We have rerun this analysis for the fruit-fly connectome using the
standard Kuramoto equation for different couplings, i.e. near the
Hopf synchronization transition discussed in~\cite{Flycikk}.

\subsection{Crackling noise analysis}

\begin{figure}[!htbp]
	\centering
	\includegraphics[width=0.46\textwidth]{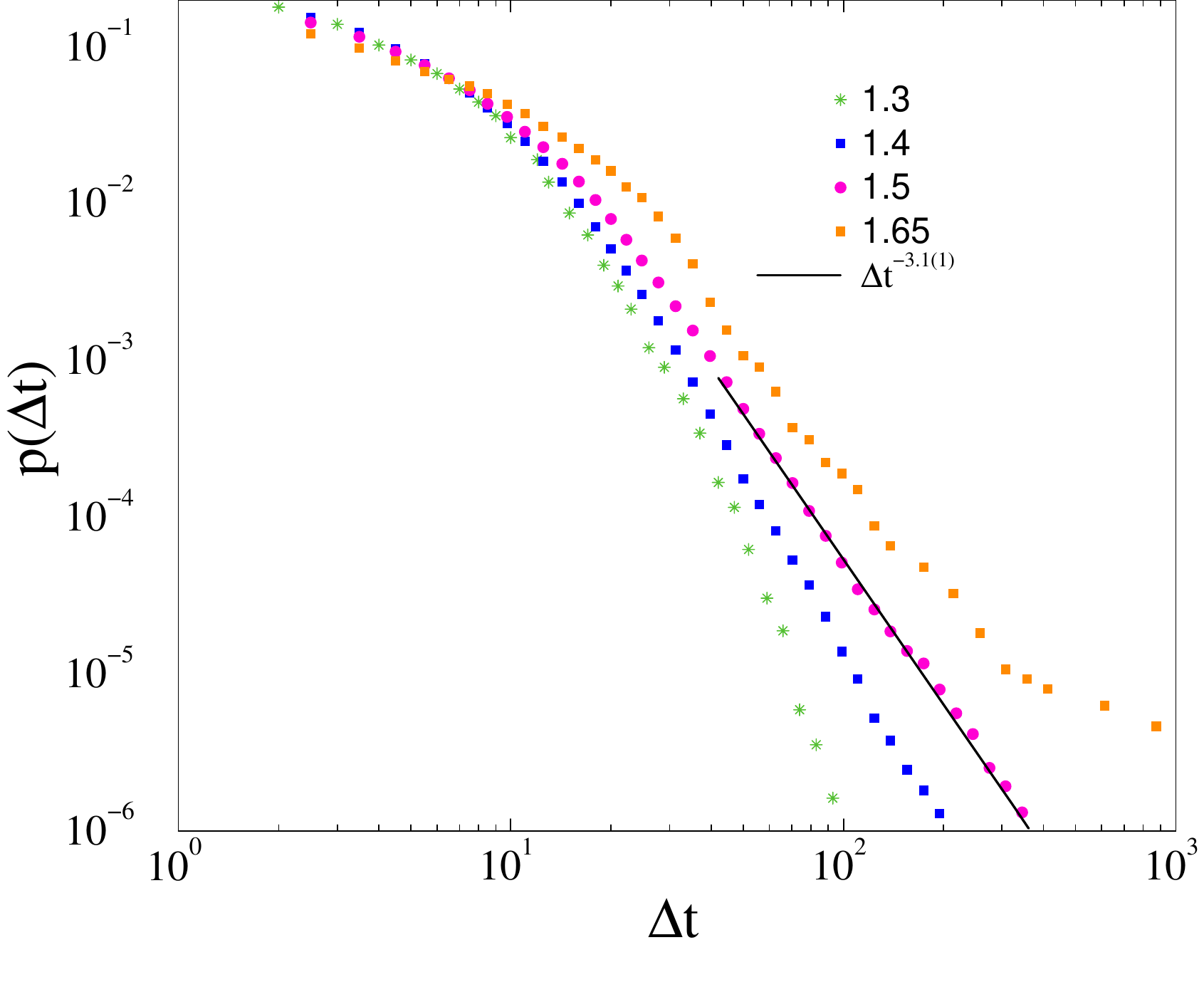}
	\caption{Avalanche duration distributions on the fruit-fly
		connectome without force for $K$ different couplings.
		\label{elo-O.fly}}
\end{figure}

Earlier mean-field type of phase transition was found at 
$K_c\simeq 1.7(2)$. As we can see on Fig.~\ref{elo-O.fly} the
crackling noise duration analysis results in faster than PL 
decays of $p(\Delta t)$ for $K < 1.4$ and an inflection point
with up veering decays for $K \ge 1.65$ couplings. 
At $K=1.5$ we can observe a PDF, with PL decay for $30 < t < 300$, 
which can be fitted by the exponent $\alpha=3.03(3)$.

As in~\cite{Flycikk} we do not find an extended scaling region with  
non-universal exponents suggesting a GP. So, the crackling noise
exponent, presumably the mean-field class exponent of the
Hopf transition, describing the resting state, should be
this value. This is a rather large exponent and is difficult to
reproduce by simulations, because large systems are needed
to see the scaling region before an exponential cutoff.
We assume that this was not seen in~\cite{PhysRevResearch.3.023224},
where $N=500$ nodes were used. Another reason might be that 
in~\cite{PhysRevResearch.3.023224} an annealed Kuramoto model 
was simulated, lacking the quenched self-frequencies.
Or perhaps because \cite{PhysRevResearch.3.023224} used
thresholds of the $\theta_i(t)$ variables and identified
avalanches by estimating the spatio-temporal size of the
activity avalanches.

But indeed the scaling region we observe is rathernarrow, 
even though we know that the Kuramoto model exhibits
a critical synchronization transition here.

\subsection{Hurst and $\beta$ analysis of local variables} \label{HBnF}

We cannot exclude the possibility that doing the avalanche
analysis on the local angles $\theta_i(t)$, would lead to a
lack of PL-s as it was claimed in ~\cite{PhysRevResearch.3.023224}.
Since identification of avalanches of local variables is a rather
difficult and ambiguous task, requiring careful binning,
to check the scaling of local phase and frequency data we performed
auto-correlation measurements and estimated the Hurst and $\beta$
exponents as before.

The "no-force" result on Figs.~\ref{fruitfly-hurst}
and ~\ref{human-hurst} show strong auto-correlations, 
indications of criticality as in the brain experiments ~\cite{Ochab2022}.
In fact the exponents are larger (close to 1), than in case of the 
Shinomoto--Kuramoto model calculations.
This suggests that the external excitation results in a less
correlated scaling behavior of the neural systems than in the
resting state. These results are in agreement with the
experimental findings of ~\cite{Ochab2022}.

\section{Conclusions}

In conclusion our numerical analysis of synchronization models
on different connectome graphs show that in the case of excitation
we can find PL scaling of duration of the crackling noise of the
activity, defined by thresholds of $R$. By solving the 
Shinomoto-Kuramoto model numerically we concluded that even without 
the additive noise we can find similar synchronization transition
as with the full Langevin equation.

The observed PL tails exhibit some dependence on the amplitude of 
the force, which may be related to GP heterogeneity effects, but 
can also arise as the consequence of quasi-critical, scaling like 
behavior reported in the discrete models of Ref.~\cite{Fosque-20}.
We estimated the extension of the synchronization transition region 
by the fluctuations of $R$ and $\Omega$ and found an extended, 
smeared transition region. This makes it difficult to define the
transition points. We attempted it in two different ways: half
values and fluctuation peaks of the order parameters in the 
steady state. The $F'_c$ from the $R$ half values are somewhat
smaller than the $F_c$ of $\Omega$-s and agree with the frequency
fluctuation peaks. While the phase fluctuations peaks were found to
be much smaller both for the FF and the human connectomes. 
This is very different from the Kuramoto Hopf transition 
results~\cite{Flycikk}.

However, $\sigma(R)$ describes a transition of the SK order parameter,
introduced for excitable systems.
In the case of the synchronized transition in FF the addition of 
force results in a quick decay of the SK order parameter like 
in the SNIC transition.
We have not reached a region, showing hybrid phase transition 
reported in~\cite{PhysRevResearch.3.023224},
possibly by the lack of strong noise. We avoided to apply
strong noise, because that makes the numerical solution less
precise or very slow.
A systematic finite-size scaling study of this transition would 
be necessary to settle this issue.

In case of initial conditions with random phase variables the $R(t)$
curves at the transition point do not show PL growth as in case 
of the  Kuramoto model, but a logarithmic growth, similar
to strong random fixed points of models of statistical
physics. 

We also investigated the local order parameters and found frustrated 
synchronization with Chimera like states, coexistence 
of synchronized and asynchronous domains. 
Performing auto-correlation analysis on the local order parameters
we found strong auto-correlation in the resting (Kuramoto) 
state at criticality and somewhat weaker ones in presence of an
external force. In the latter case the $H$ and $\beta$ exponents 
take their maximal values, where the fluctuations of $R(t)$ are 
maximal, i.e at the transition.

We also investigated the module dependence of $H$ and $\beta$ by
decomposing the connectomes via community detection algorithms.
We observed variations amongst the communities suggesting different
levels of criticality, but the identification of communities with
real brain regions is a further task to be completed.
Our simulated $H$ and $\beta$ exponents are in agreement with 
recent experimental findings~\cite{Ochab2022}.

\acknowledgments{
We thank R\'obert Juh\'asz for the useful comments.
Support from the Hungarian National Research, Development and 
Innovation Office NKFIH (K128989) is acknowledged.
Most of the numerical work was done on KIFU supercomputers of Hungary.
}

\section*{Appendix}

\begin{figure}[!htbp]
        \centering
        \includegraphics[width=0.46\textwidth]{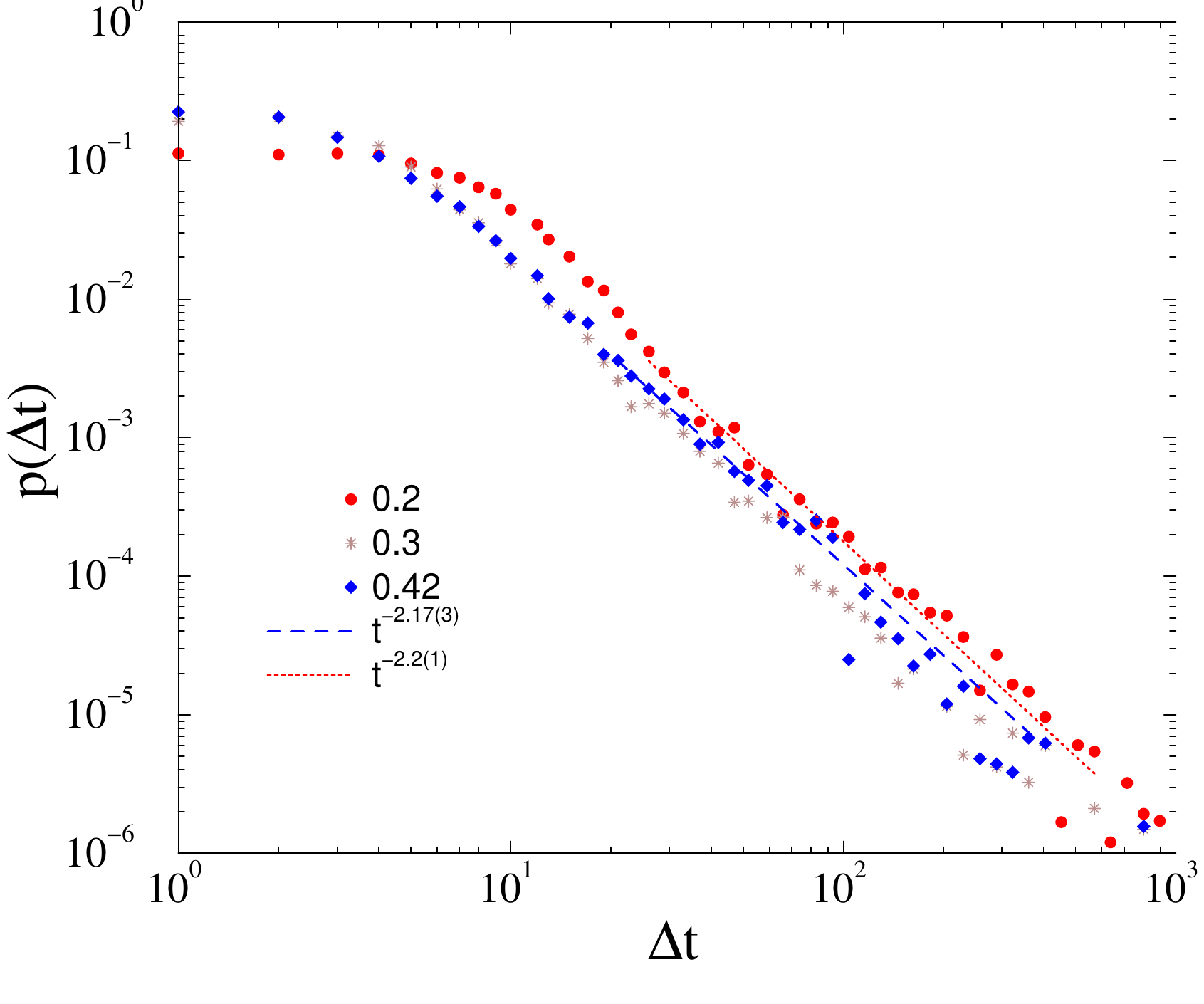}
        \caption{Avalanche duration distributions on the KKI-113
        connectome for different forces, shown by the legends and
        at $K=1$, without noise. Dashed lines are PL fits
        for $\Delta t > 20$.
        \label{elo-o.113-kurl1_F}}
\end{figure}

Here we show avalanche duration PDF-s without noise in case of the
KKI-113 connectome on Fig.~\ref{elo-o.113-kurl1_F}. One see only a
slight variation of the PL tail exponents around $-2.2$, but they are
close to the noisy case results. 

\begin{figure}[!htbp]
        \centering
        \includegraphics[width=0.44\textwidth]{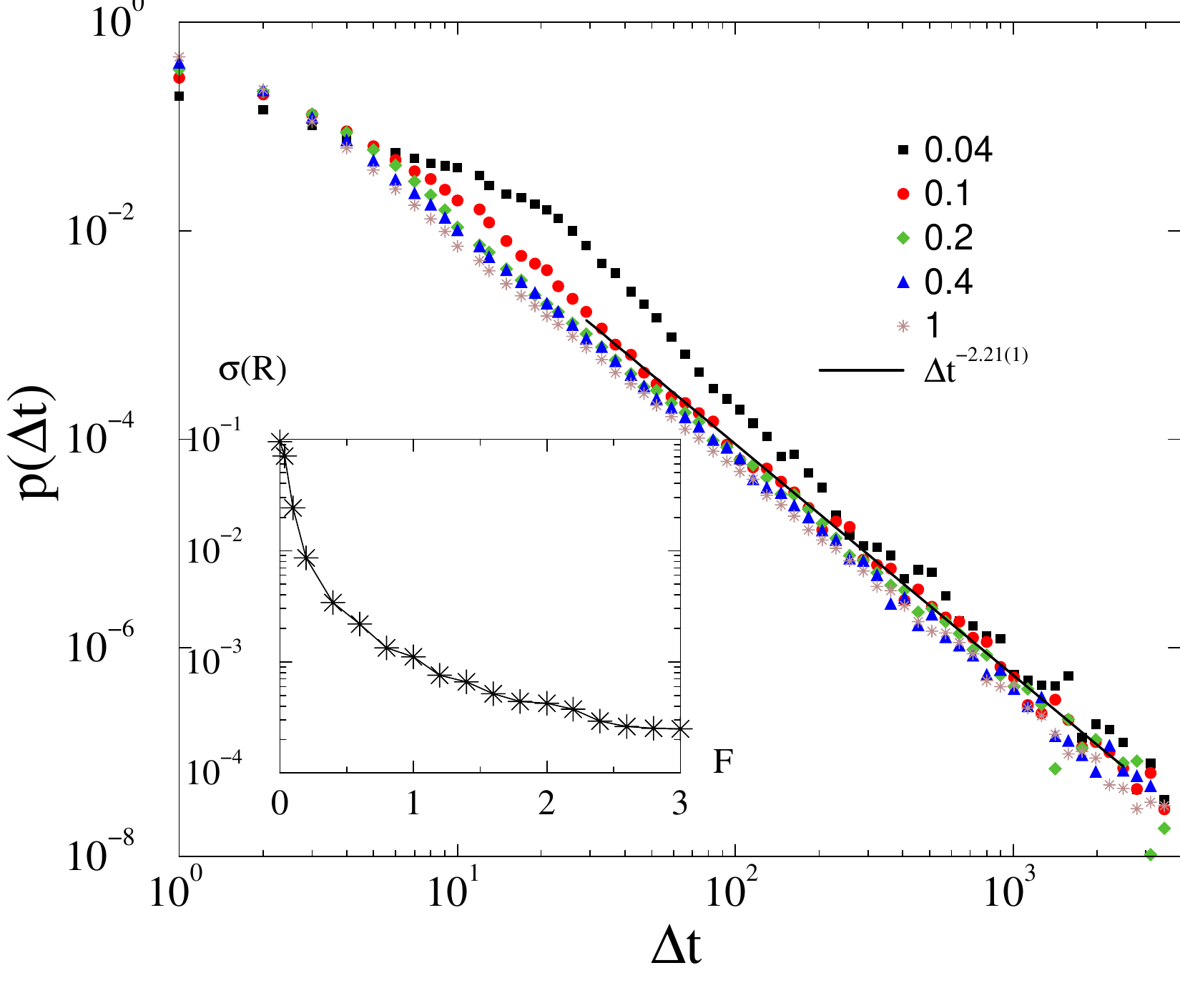}
        \caption{Avalanche duration distributions on the fruit-fly
        connectome for different forces, shown by the legends
        and at $K=2$, $\epsilon=0.01$. Dashed lines are PL fits
        for $\Delta t > 100$. The inset shows $\sigma(R)$ by
        increasing $F$.
        \label{elo-o.flyF_l2_N0.01}}
\end{figure}

Similarly, in case of the FF with the application of force in the
synchronized phase, i.e. $K=2$ the PL tails fitted for $t>20$ do
not differ to much, they can be characterized by an exponent $-2.21(1)$
as one can see on Fig.~\ref{elo-o.flyF_l2_N0.01}.
The inset shows the rapid drop of the SK order parameter as the function of
the force and the maximum both of $\sigma(R)$, $\sigma(\Omega)$ are
at $F \simeq 0$

%
\bibliography{bib}

\end{document}